\documentclass[12pt]{iopart}

\usepackage{cite}
\usepackage{hyperref}
\expandafter\let\csname equation*\endcsname\relax
\usepackage{lineno}
\expandafter\let\csname endequation*\endcsname\relax
\usepackage{amsmath}
\usepackage{xcolor}
\usepackage{graphicx}
\usepackage{enumerate}

\usepackage{caption}

\begin{document}

\title{A catastrophic approach to designing interacting hysterons}

\author{Gentian Muhaxheri$^{1}$\footnote{Corresponding author: gmuhaxhe@syr.edu}, Victoria Antonetti$^{2}$ and Christian D. Santangelo$^{1}$}
\address{$^{1}$ Department of Physics, Syracuse University}
\address{$^{2}$ Division of Applied Mathematics, Brown University}

\begin{abstract}
    We present a framework for analyzing collections of interacting hysterons through the lens of catastrophe theory. By modeling hysteron dynamics as a gradient system, we show how to construct hysteron transition graphs by characterizing the fold bifurcations of the dynamical system. Transition graphs represent the sequence of hysterons switching states, providing critical insights into the collective behavior of driven disordered media. Extending this analysis to higher codimension bifurcations, such as cusp bifurcations and crossings of fold curves, allows us to map out how the topology of transition graphs changes with variations in system parameters. This approach can suggest strategies for designing metamaterials capable of encoding targeted memory and computational functionalities, but it also highlights the rapid increase of design complexity with system size, further underscoring the computational challenges of controlling large hysteretic systems.
\end{abstract}

\section{Introduction}
Driven disordered media, such as sheared amorphous solids \cite{regev2021topology,keim2020global,galloway2022relationships,mungan2025self,kumar2025self,mungan2025self2}, compressed crumpled sheets \cite{shohat2022memory,jules2022delicate}, or disordered magnets \cite{katzgraber2006hysteretic}, undergo complex sequences of transitions between multiple metastable states. 
They are often modeled as a collection of bistable elements, known as hysterons, that switch between two states based on the history of a driving field \cite{preisach1935magnetische}. Individual hysterons are often modeled as exhibiting a piecewise linear relationship between two conjugate variables, \textit{e.g.} magnetization and external field, pressure and volume \cite{muhaxheri2024bifurcations}, displacement and force \cite{van2021profusion,jules2022delicate,shohat2025geometric,liu2024controlled}, current and voltage \cite{altman2025collective}, with sharp jumps at specific critical ``switching'' fields \cite{van2021profusion, szulc2022cooperative}.

When the hysterons are independent, the transitions give rise to well-characterized collective phenomena such as return-point memory \cite{keim2019memory}. But
when hysterons interact, their transitions can exhibit much more complex behaviors, including transient memory \cite{paulsen2014multiple,keim2011generic}, multiperiodic orbits \cite{lindeman2021multiple,keim2021multiperiodic}, scrambling and avalanches \cite{van2021profusion,bense2021complex}.
While important for understanding disordered systems, networks of hysterons have also emerged as a potential framework for creating metamaterials capable of encoding memory and computation \cite{paulsen2024mechanical, bense2021complex, kwakernaak2023counting, liu2024controlled, meulblok2025path, lindeman2025generalizing, altman2025collective,ren2025dynamics,sirote2024emergent,kamp2025reprogrammable}.
A major challenge in realizing meta-materials with targeted computational and memory-storing capabilities lies in understanding how material parameters influence the system's collective behavior.

%These can be supplimented by kinematic models to make unambiguous predictions about destination states after switching \cite{paulsen2024mechanical, shohat2024geometric}. 

This raises interesting questions about how well networks of hysterons can be programmed to execute certain transitions \cite{paulsen2024mechanical}, especially since hysteron interactions lead to a proliferation of possible transitions \cite{van2021profusion}.
In fact, hysterons that involve a continuous, nonlinear relationship between two conjugate variables can sometimes lend themselves to a geometric approach rooted in the analysis of bifurcations \cite{muhaxheri2024bifurcations,meulblok2025path}. 
In this paper, we study collections of interacting hysterons through the lens of catastrophe theory, a mathematical framework developed by Ren\'e Thom to classify discontinuous transitions, or catastrophes, in a system even when the underlying parameter changes smoothly \cite{thom2018structural}. One advantage of this approach is that it allows multiple variables to be varied simultaneously, offering insight into how specific parameters affect collective behavior.

This framework has been applied to various systems, including in guiding the design of multistate machines \cite{yang2023bifurcation} and studying the dynamics of gene regulation \cite{rand2021geometry, saez2022dynamical}. However, since the results of catastrophe theory are local and insufficient to fully capture transitions between stable states, it becomes essential to incorporate information about the system’s broader dynamics \cite{rand2021geometry}. 
In this paper, we combine catastrophe theory with the dynamics of a gradient system to introduce a method for determining the transition pathways of a system of interacting hysterons under global field driving. This approach is based on identifying and characterizing fold bifurcation points. We further highlight the significance of higher codimension bifurcations in designing desired transition pathways, as well as the challenges involved in locating such points.

\section{A dissipative system of interacting hysterons}
\begin{figure*}
  \centering
  \includegraphics[width=\textwidth]{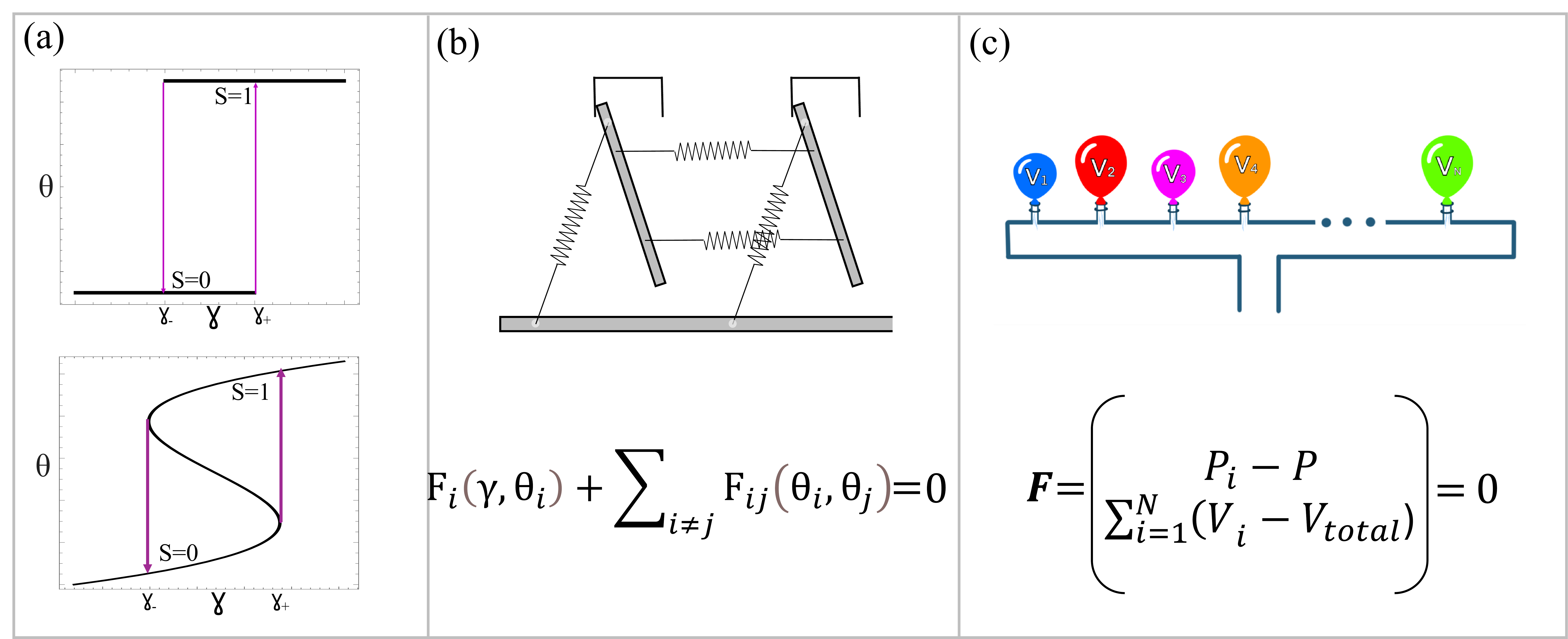}
  \captionsetup{width=\linewidth}
  \caption{
(a)  Schematic of a single hysteron showing its two
stable configurations, 0 and 1. Bottom: A hysteron with continuous nonlinear relationship between the two conjugate variables $\gamma$ and $\theta,$ which we call a continuous hysteron. Top: A binary hysteron. (b) Mechanical rotors connected to a driving field and to other hysterons through linear springs \cite{paulsen2024mechanical}. Below the schematic, the equilibrium equations for the system are shown, where $F_i$ represents applied torque on rotor $i$ from the displacement field, while $F_{ij}$ represents torque applied on hysteron $i$ from hysteron $j.$ (c) A system of connected balloons inflated under volume control \cite{muhaxheri2024bifurcations}, and the equilibrium equations for that system.}
  \label{fig:examples of hysterons}
\end{figure*}

We will focus primarily on a system of $N$ 'continuous' hysterons whose dynamics satisfy the gradient system,
\begin{equation}
\label{eq:dynamics-eq}
    \boldsymbol{\dot{\theta}}=-\nabla_{\boldsymbol{\theta}
    } V(\boldsymbol{\theta},\gamma,\boldsymbol{\Omega})\equiv \boldsymbol{F}(\boldsymbol{\theta},\gamma,\boldsymbol{\Omega}).
\end{equation}
We use $\theta_i$ as a continuous variable, whose value determines the state of hysteron $i$, as shown in the bottom of Fig. \ref{fig:examples of hysterons}(a), and $\boldsymbol{\theta}=\{\theta_1,\theta_2,\dots,\theta_N\}$ to denote the $N$-dimensional state space of the system. The $N$ hysterons are subject to a global driving field, $\gamma$, and all the other controllable parameters of the system are grouped into the vector $\boldsymbol{\Omega}.$

In order to have an example in mind, one can consider either a system of overdamped rotors joined by springs \cite{paulsen2024mechanical},  inflating balloons sharing a volume 
\cite{muhaxheri2024bifurcations}(Fig. \ref{fig:examples of hysterons}), elastic beams sharing a total displacement \cite{liu2024controlled}, or connected cylindrical origami bellows under force driving \cite{jules2022delicate}.

Though our approach is quite a bit more general, we will focus primarily on the rotor system of Ref. \cite{paulsen2024mechanical} because of its flexibility to encode interactions and the identification of each hysteron with a distinct physical element (Fig. \ref{fig:examples of hysterons}b). In that system, the values of the spring constants and the mounting positions of the springs are the controllable parameters, $\boldsymbol{\Omega}$, the orientation of the rods are the continuous state variables $\boldsymbol{\theta}$, and the position of the bar to which the rods are connected is the global driving field $\gamma$.
The equilibrium configurations are described by the fixed points of Eqs. \ref{eq:dynamics-eq}, which are found by solving the $N$ equations $\boldsymbol{F}(\boldsymbol{\theta},\gamma,\boldsymbol{\Omega})=0$ to obtain $\boldsymbol{\theta}$ for a given value of $\gamma$ and $\boldsymbol{\Omega}$. 

We assume that each hysteron interacts with the global driving field $\gamma$ through the potential,
\begin{equation}
    \label{eq:potential example}
    U_i(\theta_i,\gamma,\boldsymbol{\omega_i})=\frac{1}{4}a_i\theta_i^4+\frac{1}{3}b_i \theta_i^3+\frac{1}{2}c_i\theta_i^2 - \gamma \theta_i,
\end{equation}
where $\boldsymbol{\omega_i}=\{a_i,b_i,c_i\}$. Eq. (\ref{eq:potential example}) is the simplest polynomial that allows for two stable states and the ability to switch between them as $\gamma$ is varied, but we also expect it to qualitatively capture the behavior of a wide array of bistable elements.
We assume that element $i$ interacts with element $j$ through the potential,
\begin{equation}
    \label{eq:potential interaction}
    U_{ij}(\theta_i,\theta_j,k_{ij})=\frac{1}{2}k_{ij}(\theta_i-\theta_j)^2.
\end{equation}

The controllable system parameters are $\boldsymbol{\Omega}=\{\boldsymbol{\omega_i},\boldsymbol{K}\},$ where $\boldsymbol{K}$ includes all the interaction parameters $k_{ij}.$ We can write the total potential energy for the system as,
\begin{equation}
    \label{eq:potential total}
    V(\boldsymbol{\theta},\gamma,\boldsymbol{\Omega})=\sum_i^N U_i(\theta_i,\gamma,\boldsymbol{\omega_i}) + \frac{1}{2}\sum_{i\neq j}U_{ij}(\theta_i,\theta_j,k_{ij}).
\end{equation}
This potential allows for each hysteron to interact with any other hysteron in the system. If all the interaction constants vanish such that $k_{ij}=0,$ we obtain the Preisach model of hysteresis \cite{preisach1935magnetische}.

\section{From bifurcations to transition graphs} \label{sec:interacting hysterons as a gradient system}

The \textit{stable} fixed points of $\mathbf{F}$ represent stable states in the dissipative system Eq. (\ref{eq:dynamics-eq}). Each node in a transition graph is represented by a particular stable fixed point.
In the absence of interactions, each of these fixed points can be labeled according to the individual states of each hysteron. We say the system is in state $(0,0,\cdots, 0)$ when each hysteron is in the left-most minimum of $U_i$, and we say that a hysteron is in state $1$ whenever $\theta_i$ is in the right-most minimum of $U_i$. Even when there are interactions, this identification is usually unambiguous so long as we identify $0$ with any stable value of $\theta_i$ closer to the left-most minimum of $U_i$ and $1$ when it is closer to the right-most minimum.

\subsection{Obtaining transition graphs from fold bifurcations}
\label{sec:fold to t-graph}

As the global driving field $\gamma$ changes, the system transitions from one locally stable state to another. These transitions are typically represented by a directed graph whose edges denote the transitions, at critical values of $\gamma$, between stable fixed points.
In the absence of stochasticity, the system can only leave a state if its corresponding stable fixed point disappears as $\gamma$ changes.
For smooth functions $\mathbf{F}$, we can find the bifurcations by searching for points along which the Jacobian of $\mathbf{F}$ has a vanishing eigenvalue. In fact, generically, we should expect only one eigenvalue to vanish at a time as we are able to tune only one parameter, $\gamma$, to initiate the bifurcation. Thus, loss of stability will generically occur at a fold (or saddle-node) bifurcation: a local bifurcation in which a pair of fixed points, a saddle and a node (either a stable or unstable fixed point), collide and annihilate each other.

This forms the crux of our procedure for constructing a transition graph for a given hysteric system.

The procedure can be summarized as follows:
\begin{enumerate}

\item Identify all fold bifurcations involving stable states that occur as a function of the global driving field by solving an appropriate system of equations.

\item For each fold bifurcation, determine the associated stable state and classify it as either a creation or annihilation event, depending on whether the stable state is being generated or annihilated as the driving field increases.

\item Use the system dynamics and the unstable manifold of the annihilated saddle point to find the destination stable state after each annihilation.

\end{enumerate}
In the following subsections, we describe each of these steps in more detail.

\subsection{Finding and characterizing fold bifurcation points}
\label{sec:finding fold bifurcation points}

The first two steps of our procedure are accomplished by invoking the {Sotomayor theorem}, which we can state as follows \cite{perko2013differential}:\\

\noindent \textit{Suppose that $\boldsymbol{F}(\boldsymbol{\theta_0},\gamma_0,\boldsymbol{\Omega})=0,$ and the $(n\times n)$ matrix $J=D\boldsymbol{F}(\boldsymbol{\theta_0},\gamma_0,\boldsymbol{\Omega}),$ where $D \boldsymbol{F}$ is used to denote the matrix of partial derivatives of the components of $\boldsymbol{F}$ with respect to the components of $\boldsymbol{\theta},$ has a simple eigenvalue $\lambda=0$ with eigenvector $\boldsymbol{v}$ and that $J^T$ has an eigenvector $\boldsymbol{w}$ corresponding to the same eigenvalue. Furthermore, suppose that $J$ has $k$ eigenvalues with negative real part and $(n-k-1)$ eigenvalues with real positive part and that the following conditions are satisfied:}
\begin{equation}
    \label{eq:Sotomayor thm}
    \boldsymbol{w^T}\boldsymbol{F}_{\gamma}(\boldsymbol{\theta_0},\gamma_0,\boldsymbol{\Omega})\neq 0,\:\:\:\:\:\:\:\:\:\:\boldsymbol{w^T}\left( D^2 \boldsymbol{F}(\boldsymbol{\theta_0},\gamma_0,\boldsymbol{\Omega}) (\boldsymbol{v},\boldsymbol{v})  \right)\neq 0.
\end{equation}
\textit{Then there is a smooth curve of equilibrium points of Eq. (\ref{eq:dynamics-eq}) in $\boldsymbol{R}^n \times \boldsymbol{R},$ passing through $(\boldsymbol{\theta_0},\gamma_0)$ and tangent to the hyperplane $\boldsymbol{R}^n \times \{\gamma_0\}.$ Depending on the signs of the expressions in Eq. (\ref{eq:Sotomayor thm}), there are no equilibrium points near $\boldsymbol{\theta_0}$ when $\gamma<\gamma_0$(or when $\gamma>\gamma_0$), and there are two equilibrium points near $\gamma_0$ when $\gamma>\gamma_0$(or when $\gamma<\gamma_0$).}\\

To find the fold bifurcation points then one has to solve the system
\begin{align}
    \label{eq:constraint equations}
    \boldsymbol{F}(\boldsymbol{\theta},\gamma,\boldsymbol{\Omega})=0\\
    \text{det} ~ J(\boldsymbol{\theta},\gamma,\boldsymbol{\Omega})=0,
    \label{eq:Jacobian}
\end{align}
and ensure that those solutions satisfy the conditions given by Eq. (\ref{eq:Sotomayor thm}).
Sotomayor's theorem establishes that for $C^{\infty}$ vector fields with one parameter and a fixed point having one $0$ eigenvalue, fold bifurcations are generic in the sense that any such vector field can be perturbed to a fold bifurcation point \cite{perko2013differential}.
In contrast, in addition to Eqs. \ref{eq:constraint equations} and \ref{eq:Jacobian} being satisfied, the exchange-of-stability bifurcation arises when the first inequality in Eq. \ref{eq:Sotomayor thm} is also not satisfied, whereas the pitchfork bifurcation occurs when neither inequality is satisfied, rendering these bifurcations non-generic. Consequently, transitions between different states under global field driving are expected to occur through fold bifurcations.

The Sotomayor theorem not only identifies the fold bifurcations as the generic pathway for loss of stability, but also classifies them as either creation or annihilation events, depending on the direction in which the global driving field is varied. An annihilation event with increasing $\gamma$ is a creation event with decreasing $\gamma$. The sign of the evaluated second expression in Eq. \ref{eq:Sotomayor thm} differentiates between a creation and an annihilation event. This is the second step in our procedure for constructing a transition graph. 
To complete the procedure, we finally use the system dynamics to obtain the missing information about the transitions between states.

\subsection{Using the system dynamics: the escape route}
\label{sec:escape route}

The last step in our procedure to construct the transition graph is to determine the destination states. 
Since only one eigenvalue vanishes at a fold bifurcation, we use the fact that the corresponding saddle has index $1$, indicating it has only one unstable direction.  This unstable direction provides information about how the system leaves after loss of stability \cite{rand2021geometry}.
The process is illustrated in Fig. \ref{fig:obtaining transition graph}b. As the fold bifurcation is reached, the vanishing of an eigenvalue of $\nabla \mathbf{F}$ that defines the escape route is determined by the unstable manifold of the associated saddle point \cite{rand2021geometry,saez2022dynamical}.  The unstable manifold defines the flow lines in configuration space that form the path the system follows immediately after a stable node has been annihilated. Therefore, it also determines the destination node in the transition graph. Practically, we find the destination node by integrating the dynamics near the critical global field, $\gamma_{cr}$, at which the stable node is annihilated. We set $\gamma = \gamma_{cr} + \epsilon,$ where $\epsilon$ is a small number ($\epsilon \sim 10^{-3})$ chosen so that the dynamics still reflect the local behavior near the bifurcation. In our examples, we consider strictly overdamped dynamics, where the system cannot overshoot any energy minimum, regardless of how shallow it is. However, previous studies \cite{jin2025dynamic} have shown that even a small amount of inertia can play a significant role in determining the nature of the system’s transitions.

\subsection{Example: A system of two hysterons}
\label{sec: example}

\begin{figure*}
  \centering
  \includegraphics[width=\textwidth]{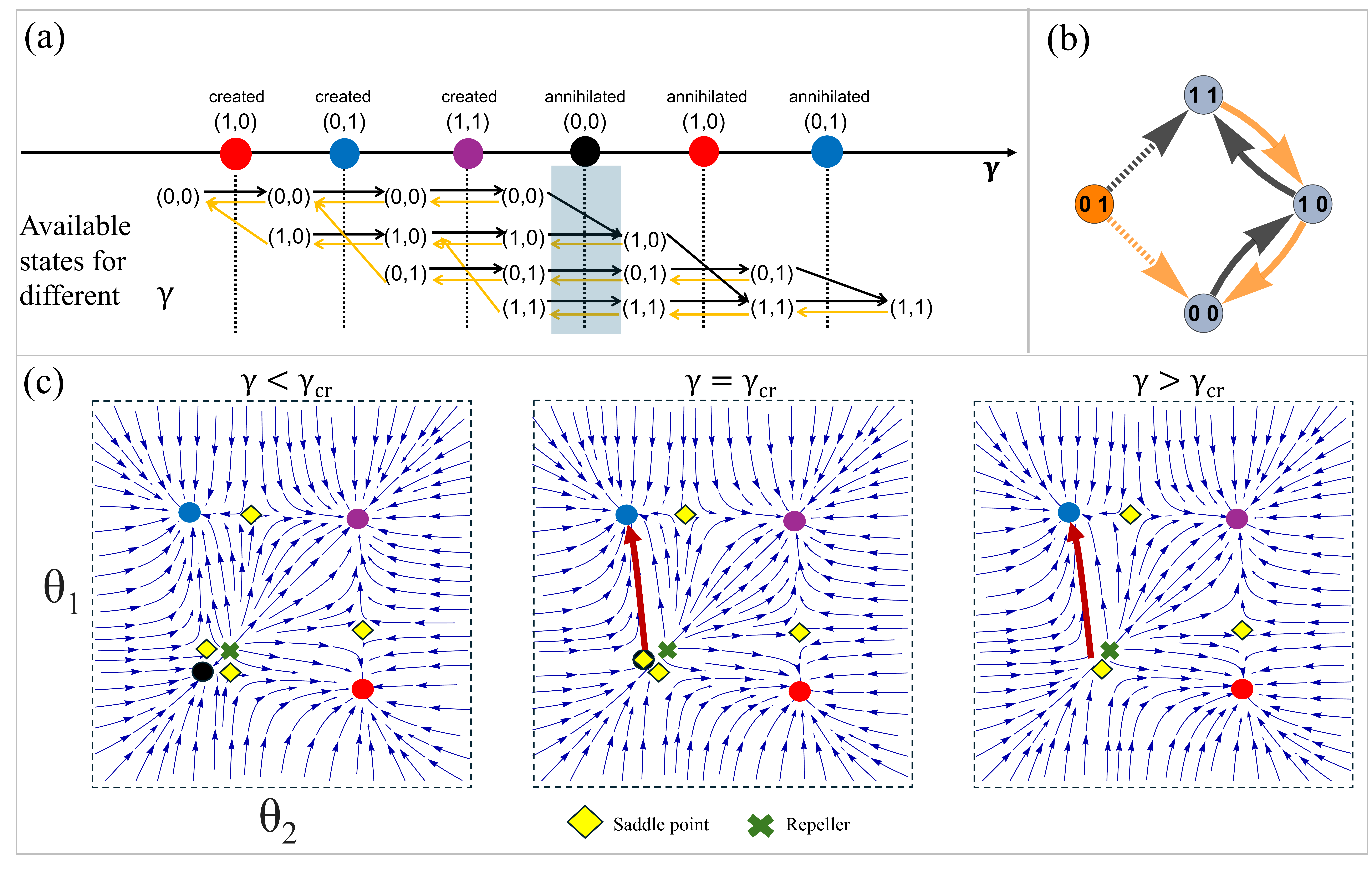}
\captionsetup{width=\linewidth}
  \caption{Obtaining the transition graph for an example of two hysterons. (a) Each fold bifurcation point with varying $\gamma$ is shown. Creation and annihilation label on the points assumes increasing $\gamma,$ while that labeling is inverted when $\gamma$ is decreasing. The black (orange) arrows represent transitions with increasing (decreasing) $\gamma.$ The figure shows that the system stays in a state until it is annihilated through a fold bifurcation point, in which case it transitions to a different state - that is shown through the diagonal arrows. (b) The transition graph for this system is shown. The dashed lines represent transitions out of a 'Garden of Eden' state, which is a state that cannot be accessed by varying global field $\gamma.$ The 'Garden of Eden' state is denoted by an orange vertex. (c) The dynamics of the system are shown for the $\gamma$ domain highlighted in part (a), explicitly showing the formation of the escape route, denoted by the red arrow, during the fold bifurcation involving the state $(0,0).$ The escape route directs the system to transition from $(0,0)$ to $(1,0).$   }
    \label{fig:obtaining transition graph}
\end{figure*}

To illustrate the method, we present a quantitatively simple, but quite general, dynamical system for two interacting hysterons.
We write the potential energy for a system of two interacting hysterons as,
\begin{equation}
    \label{potential two hysteron example}
    V(\boldsymbol{\theta},\gamma,\boldsymbol{\Omega})= U_1(\theta_1,\gamma,\boldsymbol{\omega_1})+U_2(\theta_2,\gamma,\boldsymbol{\omega_2})+\frac{1}{2}U_{12}(\theta_1,\theta_2,k_{12})+\frac{1}{2}U_{21}(\theta_2,\theta_1,k_{21}).
\end{equation}
The equilibrium equations are given by
\begin{equation}
    \label{eq:equilibrium eq for exmaple}
\boldsymbol{F}=-\begin{Bmatrix}
a_1 \theta_1^3+b_1 \theta_1^2 +c_1 \theta_1 - \gamma+ k_{12}(\theta_1-\theta_2)\\
a_2 \theta_2^3+b_2 \theta_2^2+c_2 \theta_2 - \gamma+ k_{12}(\theta_2-\theta_1)
\end{Bmatrix}=0,
\end{equation}
where we have assumed that $k_{12}=k_{21}.$ We set the parameters $\{a_1,b_1,c_1,a_2,b_2,c_2,k_{12}\}=\{4.4,1,-5,2,1,-3.5,-1.5\},$ and
solve Eqs. \ref{eq:constraint equations} and \ref{eq:Jacobian} for this system. We then only keep the fold bifurcation points involving stable states, characterize them using the conditions given by Sotomayor's theorem, and plot the values of $\gamma$ at which the fold bifurcation points occur, as shown in Fig. \ref{fig:obtaining transition graph}(a). This allows us to see the available stable states for the system at different regions of the driving field $\gamma.$ Furthermore, we know that when the system is already at a stable state, it stays there until the state annihilates through a fold bifurcation, and that is represented through the horizontal arrows pointing between the same state. When a state is annihilated, the system finds another destination stable state in which to land, and we find that information through finding the escape route, as we discussed in the previous section. Those transitions are represented by the non-horizontal arrows in Fig. \ref{fig:obtaining transition graph}(a). In Fig. \ref{fig:obtaining transition graph}(b), we present snapshots of the dynamical flow lines in configuration space, corresponding to values of $\gamma$ just before and just after the annihilation of the state $(0,0).$ At the critical value $\gamma=\gamma_{cr},$ stable state $(0,0)$ collides with the corresponding saddle node which forms the escape route indicated by the red arrow. Once the state is annihilated, the flow lines direct the system to settle in the adjacent stable state $(1,0).$ A similar process occurs with each state annihilation, and the result is shown in the transition graph in Fig. \ref{fig:obtaining transition graph}(c).

From the \textit{expanded bifurcation diagram} shown in Fig. \ref{fig:obtaining transition graph}, not only can we find the transition graph, but we can also see all the available states for different values of $\gamma,$ including the \textit{Garden of Eden} states, which are states that cannot be reached by global field driving. The system has to either start at that state or be forced into it for it to be part of the pathway. The state $(0,1)$ is a \textit{Garden of Eden} state in the example shown in Fig. \ref{fig:obtaining transition graph} since the system never transitions into that state, but it can start there since it is a stable state for a particular domain of $\gamma$ values. One advantage of this expanded bifurcation diagram is that it remains one dimensional, no matter the number of hysterons, as long as there is only one varying parameter in the system, which in this case is the global driving field $\gamma$.

\section{\textbf{Modifying transition graphs through bifurcations of higher co-dimension}} \label{sec: modifying t-graphs}

In Sec. \ref{sec:fold to t-graph}, we demonstrated how fold bifurcations and the system’s dynamics can be used to construct transition graphs.
Mapping the bifurcations also provides new insight into how the transition graphs themselves can be modified by varying the system parameters, $\boldsymbol{\Omega}$. This naturally points to bifurcations of higher codimension, where the codimension is equal to the number of parameters that are varied. We first focus on bifurcations of codimension $2$, and we will later see that the ideas extend naturally to higher codimensions.

When an extra parameter is varied, the bifurcation diagram features fold bifurcation curves rather than isolated points. These curves can interact in two important ways: they can meet tangentially at a cusp bifurcation point, or they can intersect at a crossing. Each of these alter the expanded bifurcation diagram and, in turn, can change the topology of the transition graph.

\subsection{Cusp bifurcations}
\label{sec: cusp bifurcations}

There exist two types of codimension 2 cusp bifurcations: the standard cusp and the dual cusp. In a standard cusp, two fold bifurcation curves, each corresponding to a different state, converge tangentially at a single point. Essentially, the standard cusp point describes the creation (or annihilation) of a pair consisting of a stable point and a saddle, connected to an existing stable state \cite{saez2022dynamical}. Consequently, moving across the standard cusp bifurcation in parameter space determines whether the corresponding state is included or completely excluded from the transition graph. In contrast, a dual cusp occurs when two fold bifurcation curves associated with the same state meet tangentially. In a dual cusp, two saddle points and a stable node coalesce at a single point in state space. Thus, as one moves across a dual cusp in parameter space, which saddle point annihilates the stable node changes, and this, in turn, alters the system’s escape route and, consequently, the transition graph. Identifying and understanding dual cusp bifurcations is essential for designing systems with targeted transition pathways.

\begin{figure*}
  \centering
  \includegraphics[width=\textwidth]{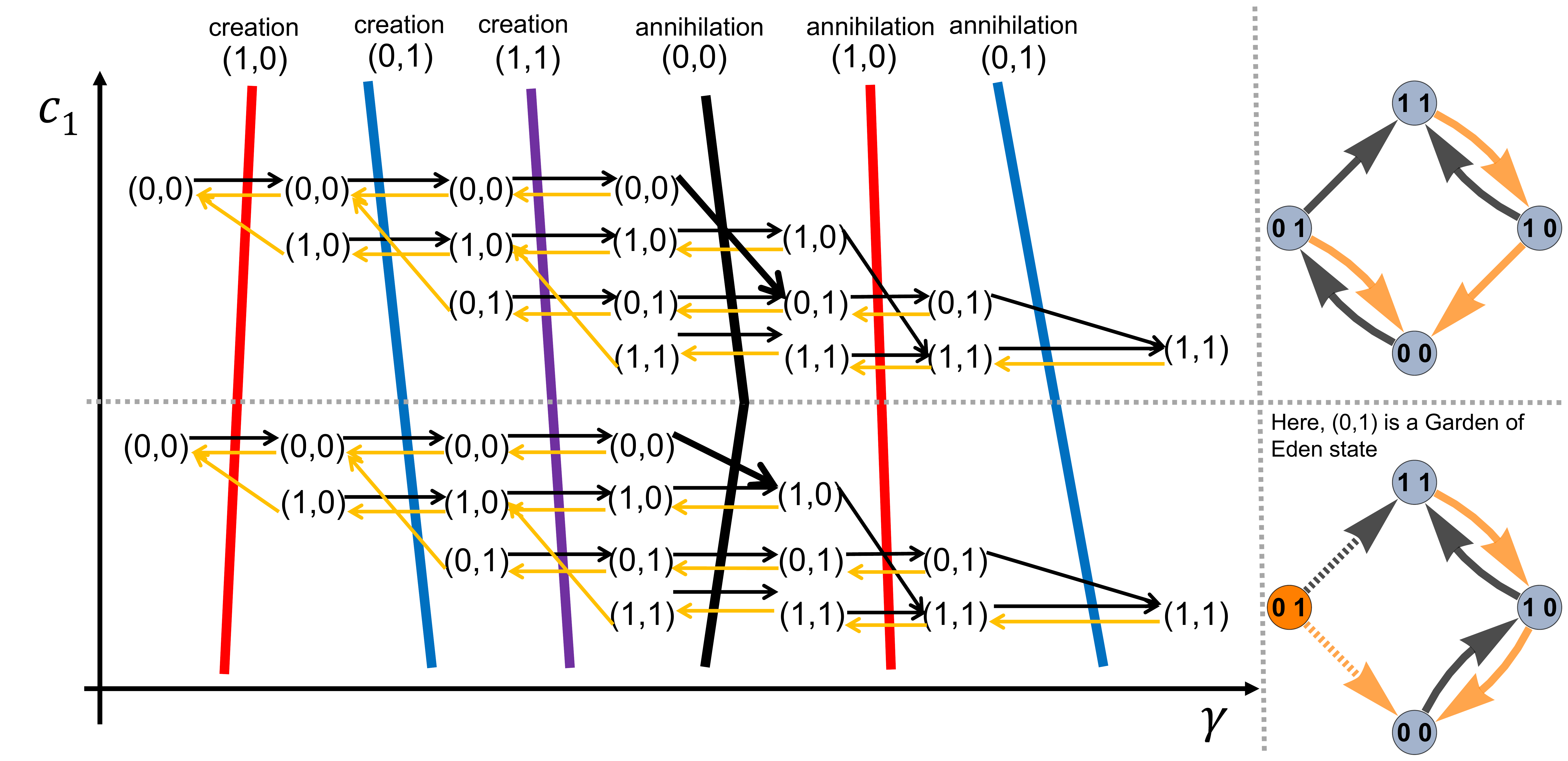}
  \captionsetup{width=\linewidth}
  \caption{
The expanded bifurcation diagram for the example in Sec. \ref{sec: example}, with varying parameter $c_1$ showing a dual cusp bifurcation involving the state $(0,0)$ which separates the diagram into two domains of the parameter $c_1,$ with each domain showing a different transition graph topology. In the transition graphs, an orange vertex denotes a 'Garden of Eden' state.}
  \label{fig:cusp bifurcation}
\end{figure*}

Fig. \ref{fig:cusp bifurcation} illustrates how the expanded bifurcation diagram from the two-hysteron example in Sec. \ref{sec: example} changes by varying parameter $c_1$. This parameter controls the shallowness of the potential minima, thereby leading to the disappearance or creation of one of the minima. The expanded bifurcation diagram includes a dual cusp bifurcation point where two fold bifurcation curves that annihilate the state $(0,0)$ meet to form a kink.
%At the dual cusp. the escape route associated with this state changes, leading to a different transition graph topology on either side of the dual cusp, consistent with the expected behavior. 

To find these cusp bifurcations generally, we first expand the dynamics of the system near a bifurcation point,
\begin{equation}
    \label{eq:expanded dynamics}
    \boldsymbol{\dot{\theta}}=J \boldsymbol{\theta}+\frac{1}{2}\boldsymbol{B}(\boldsymbol{\theta},\boldsymbol{\theta})+\frac{1}{6}\boldsymbol{C}(\boldsymbol{\theta},\boldsymbol{\theta},\boldsymbol{\theta})+\dots,
\end{equation}
where $\boldsymbol{B}, \boldsymbol{C}$ are the bilinear and trilinear forms respectively. The projection of $\boldsymbol{\theta}$ onto the center manifold, $z=\boldsymbol{w}\cdot \boldsymbol{\theta}$, has dynamics given by,
\begin{equation}
    \label{eq:center manifold dynamics}
    \dot{z}=q_2 z^2+q_3 z^3+\dots,
\end{equation}
where $q_2=(1/2)\:\boldsymbol{w}\cdot \boldsymbol{B}(\boldsymbol{v},\boldsymbol{v}),$ and, as before, $\boldsymbol{v}$ and $\boldsymbol{w}$ denote the right and the left null space vectors of the Jacobian respectively. In a similar fashion, we can write $q_3,q_4$ and the other higher-order coefficients as well. The expression for these coefficients up to fourth order are given in the SI of Ref. \cite{yang2023bifurcation}. When $q_2=0,$ $q_3\neq0,$ and Eqs. (\ref{eq:constraint equations}, \ref{eq:Jacobian}) are satisfied, the system is at a codimension-$2$ cusp bifurcation. If $q_3$ also vanishes but $q_4 \neq 0$, it is at a codimension-$3$ cusp bifurcation (sometimes called a swallowtail bifurcation).

\subsection{Crossings of fold bifurcation curves}

When two fold bifurcation curves cross transversally at a point in parameter space, that is also a bifurcation of codimension $2$. At that point in parameter space, there occur two fold bifurcations simultaneously but they correspond to two different points in state space. Similar to the cusp bifurcation, a crossing bifurcation separates the parameter space into two domains of the varying parameter, but unlike the dual cusp bifurcation, a crossing of fold curves does not always change the topology of the transition graph. However, it always changes the number of stable states available for a given driving field.
%\textcolor{blue}{[can you be more explicit as to why 4 regions or how to identify that it is four regions intuitively?]} \textcolor{red}{[I changed the sentence to talk about the splitting of parameter space into two domains instead, since that's what matters in the t-graph changes]}

\begin{figure*}
  \centering
  \includegraphics[width=\textwidth]{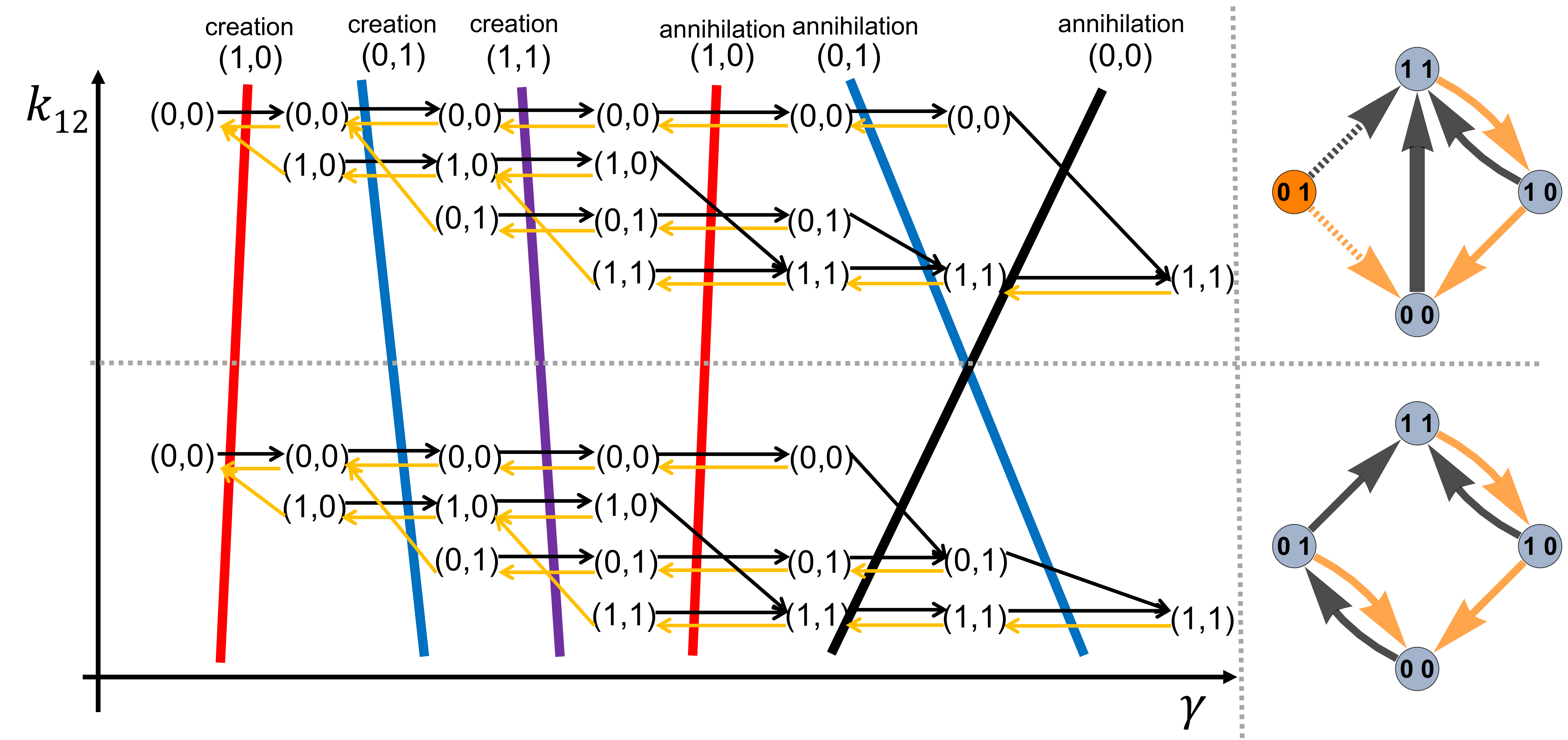}
  \captionsetup{width=\linewidth}
  \caption{
The expanded bifurcation diagram for the example in Sec. \ref{sec: example}, with varying interaction parameter, $k_{12},$ showing a crossing of fold bifurcation curves involving the states $(0,0)$ and $(0,1)$ which separates the diagram into two domains of the parameter $k_{12}.$ In the lower domain, when the state $(0,0)$ annihilates, it can transition to state $(0,1),$ but that is not possible in the upper domain after the crossing of the curves. The state $(0,0)$ can only transition to state $(1,1),$ thus causing an avalanche, and turning state $(0,1)$ into a Garden of Eden state.}
  \label{fig:crossing}
\end{figure*}

Crossings play an important role in enabling avalanches, although they do not always lead to one, nor are they the only mechanism by which avalanches can arise. An example for two hysterons is shown in the expanded bifurcation diagram in Fig. \ref{fig:crossing}, which depicts a crossing of fold curves when varying parameter $k_{12}$ that controls the hysteron interaction. There, the curve corresponding to state $(0,0)$ crosses the curve corresponding to state $(1,0)$, so when the state $(0,0)$ is annihilated, the system can no longer transition into the state $(1,0)$. Instead, it must transition to the only other available stable state, $(1,1).$ It is the simultaneous switching of two hysteron states that makes this an avalanche.

To generally find all the crossing points of a system, we have to solve for
\begin{gather}
    \label{eq:crossing}
    \boldsymbol{F}(\boldsymbol{\theta_0},\gamma_0,\boldsymbol{\Omega})=0; \:\:\: \boldsymbol{F}(\boldsymbol{\theta_1},\gamma_0,\boldsymbol{\Omega})=0\nonumber\\
     \text{det}(J(\boldsymbol{\theta_0},\gamma_0,\boldsymbol{\Omega}))=0;\:\:\:\text{det}(J(\boldsymbol{\theta_1},\gamma_0,\boldsymbol{\Omega}))=0.
\end{gather}
This constitutes a system of $2N+2$ equations, which allows us to find $\boldsymbol{\theta_0}, \boldsymbol{\theta_1}, \gamma,$ and one of the parameters in $\boldsymbol{\Omega}$.

\begin{figure*}
  \centering
  \includegraphics[width=\textwidth]{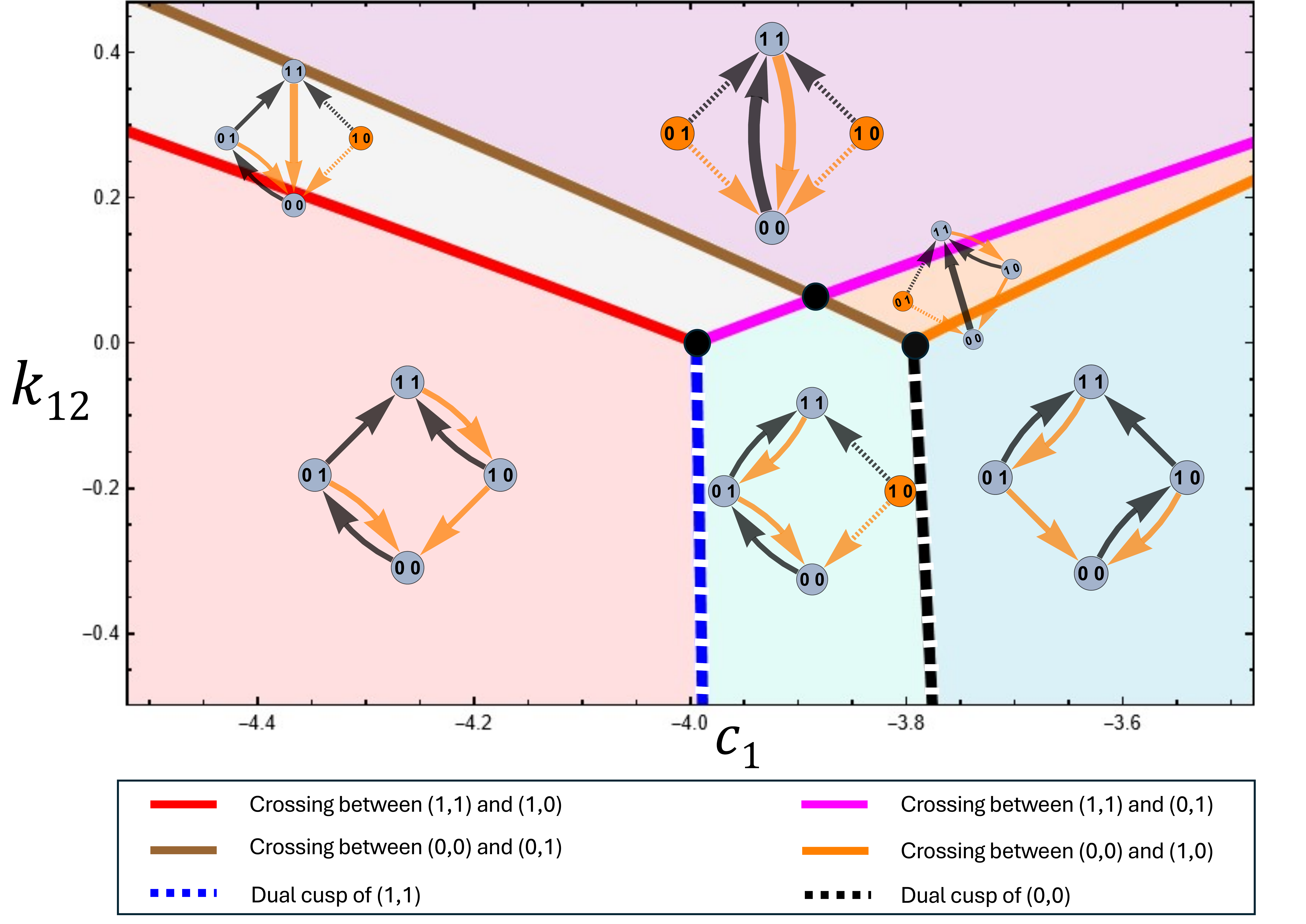}
  \captionsetup{width=\linewidth}
  \caption{The reduced two-dimensional parameter space that includes the varying parameters $c_1$ and $k_{12}$ is shown, together with curves that represent codimension $2$ bifurcations. Each point in the dashed curves represents a dual cusp bifurcation point while each point in the undashed curves represents a crossing of fold bifurcation curves. Essentially, each point in any of these curves represents a codimension $2$ bifurcation point, and the points at which these curves meet represent bifurcations of codimension $3,$ represented by the black dots in the figure. Dashed arrows in the transition graphs are used for Garden of Eden states, and thicker arrows indicate an avalanche.}
  \label{fig:graphsofgraphs}
\end{figure*}

\subsection{Bifurcations of codimension $>2$}

So far, we have shown that transition graphs can be constructed using codimension $1$ bifurcations, and that they can be modified through codimension $2$ bifurcations. This naturally leads to the consideration of bifurcations of codimension higher than $2$, which can be used to describe changes in transition graphs with additional parameters. To better understand the importance of these bifurcations, we choose to vary two parameters, $c_1$ and $k_{12}$, in addition to the driving field $\gamma$ for the example we discussed in Sec. \ref{sec: example}. Now, with three varying parameters, instead of obtaining isolated cusp and crossing points, we obtain curves of these codimension $2$ bifurcations. Since transition graph topology can change by varying parameters that are not the driving field, we choose to plot the cusp and crossing curves into the two-dimensional parameter space made up of $c_1$ and $k_{12}$ as seen in Fig. \ref{fig:graphsofgraphs}. There, the parameter space is partitioned by the curves of codimension $2$ bifurcations, with dashed curves representing dual cusps while undashed curves representing crossing of fold bifurcation curves. Neighboring transition graphs are separated by those codimension $2$ bifurcation curves and they can differ from one another by only one different edge or one vertex. We can think of going from a transition graph to another desired graph one edge or vertex change at a time, and understanding how certain system parameters affect the collective is then crucial to be able to target certain transition graphs. As depicted by the black dots in Fig. \ref{fig:graphsofgraphs}, the codimension $2$ bifurcation curves themselves can meet or cross with each other to form codimension $3$ bifurcation points, which suggests that finding these points directly can guide us to values of the parameters around which multiple different transition graphs can be accessed. Whereas two codimension $2$ bifurcation curves may intersect in parameter space while corresponding to distinct points in state space, the situation is reversed when such curves merge instead. In the latter case, the merging occurs at a single point in state space, which constitutes a codimension $3$ bifurcation point where the quantities $q_2$ and $q_3$ from Eq. \ref{eq:center manifold dynamics} are equal to zero. Obtaining the equations to solve for these points is straightforward, but solving them for an increasing number of hysterons becomes very difficult. However, visualizing the bifurcation curves in parameter space can serve as a tool to understand the way in which different system parameters affect the behavior of the collection of hysterons and, in turn, help reach desired pathways for the system of hysterons. 

\begin{figure*}
  \centering
  \includegraphics[width=\textwidth]{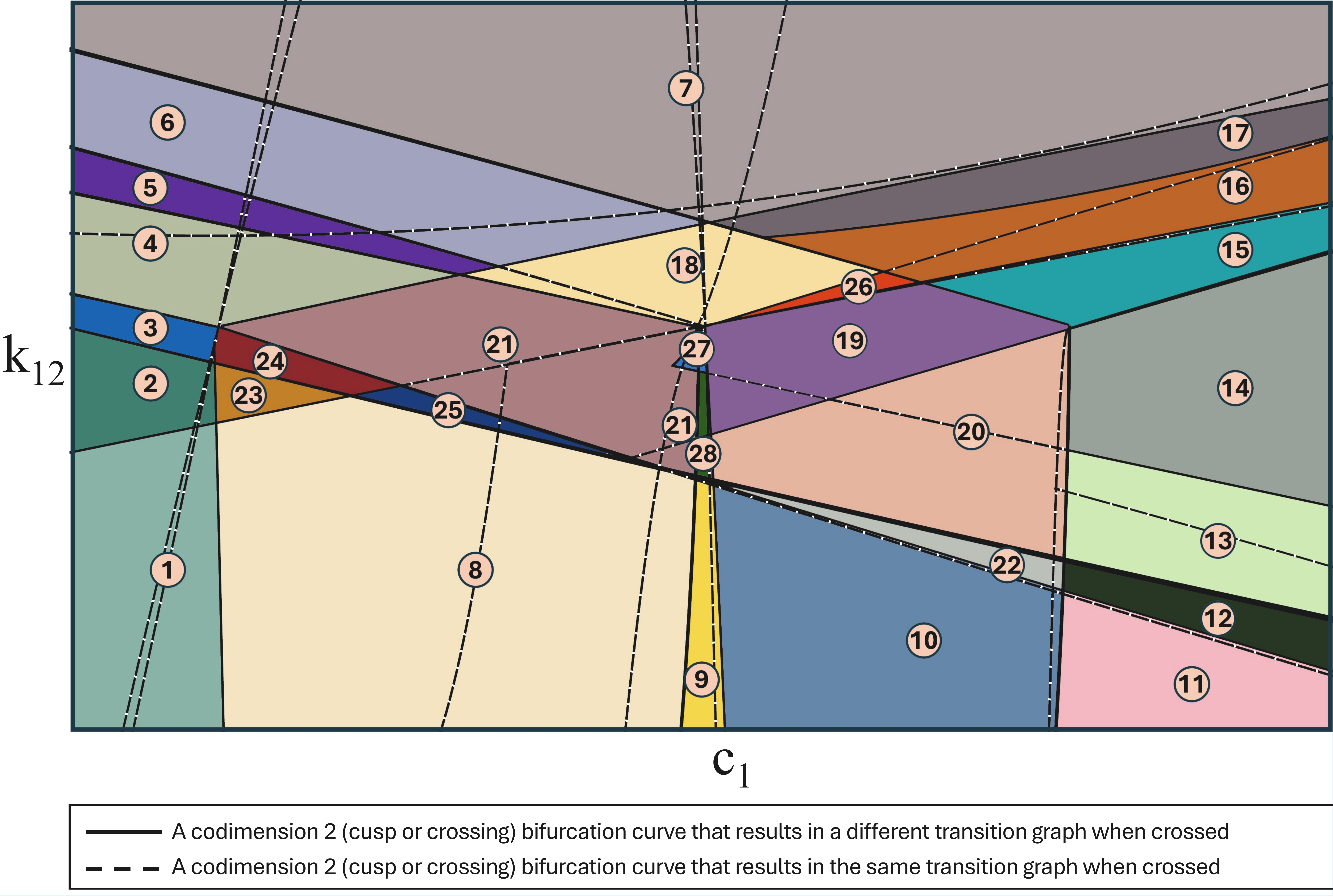}
  \captionsetup{width=\linewidth}
  \caption{The reduced two-dimensional parameter space that includes the varying parameters $c_1$ and $k_{12}$ is shown, together with curves that represent codimension $2$ bifurcations. Dashed curves represent bifurcations that do not result in a different transition graph when crosses, while undashed curves represent those that result in a different transition graph when crossed. The numbers represent which transition graph belongs to the specific region, and the numbered transition graphs are shown in Fig. \ref{fig:legend_of_final_graph}.}
  \label{fig:final_graph}
\end{figure*}

\begin{figure*}
  \centering
  \includegraphics[width=1\textwidth]{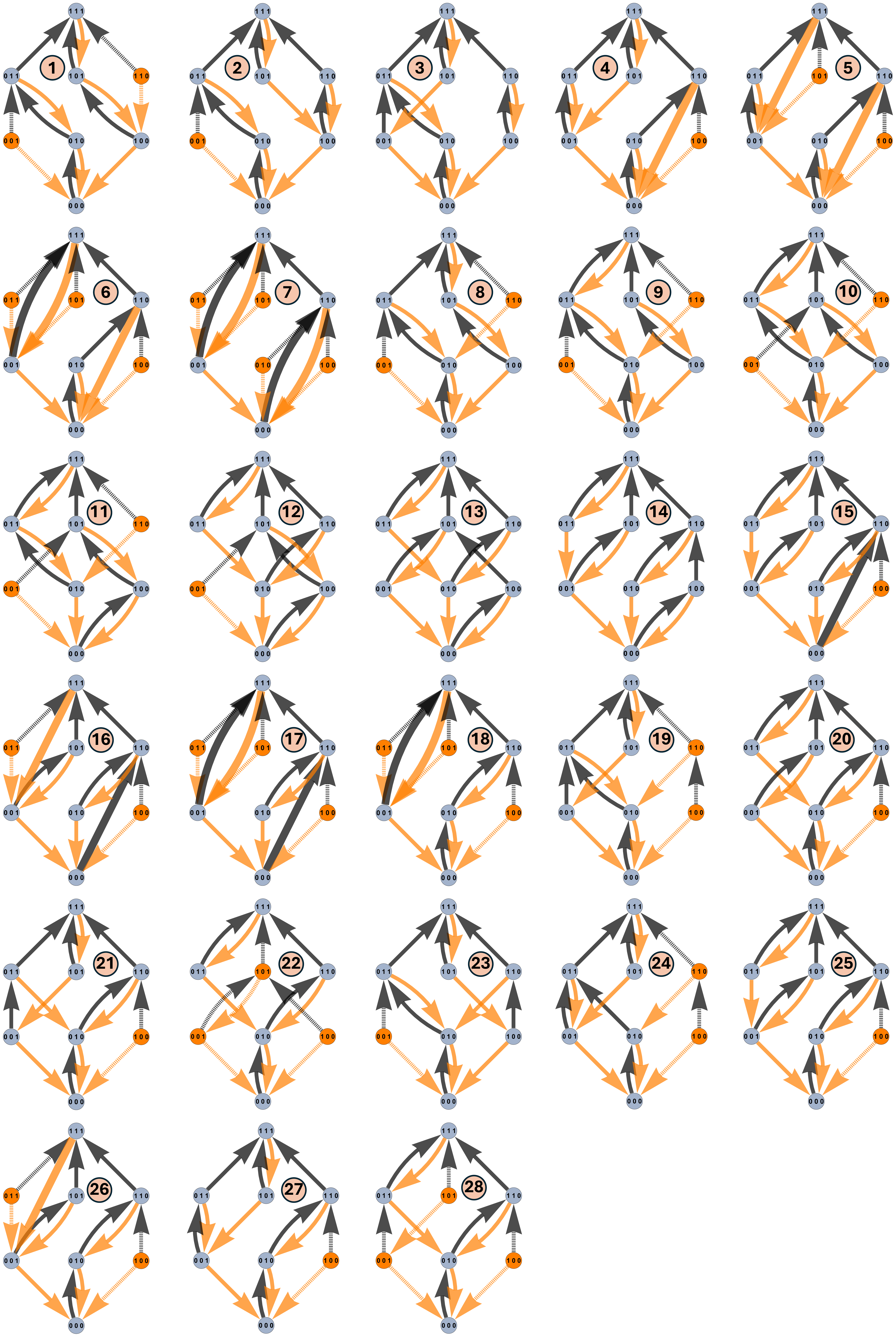}
  \captionsetup{width=\linewidth}
  \caption{Legend of transition graphs for Fig. \ref{fig:final_graph}}
  \label{fig:legend_of_final_graph}
\end{figure*}

A diagram such as the one shown in Fig. \ref{fig:graphsofgraphs} can always be drawn for any number of hysterons, as long as we vary two of the controllable parameters in the system in addition to the driving field. That's due to the fact that the transition graph can only change by crossing a cusp or a crossing bifurcation. Each distinct region in parameter space corresponds to a transition graph, though not necessarily a unique one, and the specific path taken to reach that region does not affect the resulting graph. That is, the transition graph is uniquely determined by the system parameters, and different paths through parameter space cannot lead to different graphs for the same parameter values. To illustrate this idea more concretely and highlight the growing complexity of interacting hysterons, we consider a system of three hysterons. We fix all the system parameters except $c_1$ and $k_{12}$ and draw the curves representing codimension $2$ bifurcations including cusp and crossings of fold curves (Fig. \ref{fig:final_graph}). The two-dimensional parameter space defined by $c_1$ and $k_{12}$ is now divided into more regions, though not all of them correspond to distinct transition graphs. The dashed curves represent codimension $2$ bifurcations that do not alter the topology of the transition graph when crossed, whereas the solid curves indicate bifurcation boundaries across which the transition graph topology changes. It should be noted that the shape and volume of the regions in the parameter spaces in Fig. \ref{fig:graphsofgraphs} and \ref{fig:final_graph}, as well as the transition graphs representing those regions, would be different if different parameters were chosen instead to be varied. The transition graphs belonging to the different regions in the parameter space from Fig. \ref{fig:final_graph} are shown in Fig. \ref{fig:legend_of_final_graph}. There, we can see examples of transition graphs that exhibit avalanches, which are denoted by thicker directed edges. Those can be seen in transition graphs numbered $4, 5, 6, 7, 15, 16, 17, 18, 26.$ At the same time, we can see that scrambling is present as well. For example, in transiton graph $1$ of Fig. \ref{fig:legend_of_final_graph}, there is a transition from state $(0 ,0,0)$ to state $(0,1,0),$ which suggests that the critical $\gamma$ at which hysteron $2$ transitions up is smaller than the critical $\gamma$ at which hysteron $3$ transitions up. However, the transition from state $(1,0,0)$ to state $(1,0,1)$ suggests that the opposite is true. That constitutes scrambling since $\gamma_{2+}(\boldsymbol{s}=(0,0,0)) > \gamma_{3+}(\boldsymbol{s}=(0,0,0)),$ but $\gamma_{2+}(\boldsymbol{s}=(1,0,0)) < \gamma_{3+}(\boldsymbol{s}=(1,0,0)).$

Previous work on hysteron models, where the parameter space is defined through linear inequalities, has shown that the boundaries between regions corresponding to different transition graphs are strictly linear \cite{van2021profusion,teunisse2025transition,baconnier2025dynamicselfloopsnetworkspassive}. In our case, however, the boundaries appear curved. This difference arises because the relationship between the two conjugate variables in our model is polynomial rather than piecewise linear, which naturally leads to nonlinear boundaries.

\section{Discussion}
In this work, we have presented a framework for constructing transition graphs for interacting hysterons under quasistatic driving. We do this by focusing on fold bifurcations as the generic mechanism for transitions between states, and using the unstable manifolds of the index-$1$ saddle points to determine escape routes. We further showed that codimension-$2$ bifurcations, such as cusps and crossings of fold curves, outline boundaries in parameter space where the topology of the transition graphs changes. Our method allows us to vary a small number of parameters at a time in order to identify ``nearby'' transition graphs robustly and systematically.

However, even when only two parameters are varied, the number of possible graphs appears to grow rapidly with an increasing number of hysterons. This is not altogether unexpected: the number of distinct transition graphs increases considerably with increasing number of hysterons \cite{van2021profusion,teunisse2025transition}. But Fig. \ref{fig:final_graph} shows that, even varying only two parameters and keeping the others constant, the space of transition graphs involves many subregions with varying volume. It is therefore natural to expect the system to become more sensitive to ``programming errors,'' since the areas of the regions in Fig. \ref{fig:final_graph} indicate the degree of tuning required to achieve a specific transition graph. This suggests that many possible transition graphs might become physically inaccessible due to their delicacy as the number of hysterons increases. We do not have an explanation for the relative areas of different regions in Fig. \ref{fig:final_graph}.

With some strong assumptions, however, we can estimate an upper limit for the number of regions a graph like the one shown in Fig. \ref{fig:final_graph}. A system of $N$ hysterons can have $2^N$ unique states. If we assume that we have a crossing bifurcation between every combination of stable states except the two extremes, $(0,0,\dots,0)$ and $(1,1,\dots,1)$, then we expect $\binom{2^N}{2}-1$ crossings.  If we assume that there is, at most, one dual cusp bifurcation for each state and one standard cusp bifurcation for each non-extremal state, we find that we can have $2^{N+1}-2$ total dual and standard cusp bifurcations. Adding them, we obtain $n=\binom{2^N}{2}+2^{N+1}-3=2^{2N-1}+3(2^{N-1}-1)$ bifurcations. If we plot all these codimension-$2$ bifurcation curves in the $2D$ parameter space plot and assume that they all cross each other at some point, the $2D$ parameter space would separate into $\frac{n^2+n+2}{2}$ regions. While this estimate dramatically overcounts the number of regions we do observe, it does highlight the rapid growth of complexity in the behavior of interacting hysterons, even when we only access a relatively small number of control parameters. We emphasize that this estimate does not provide an upper bound on the number of possible transition graphs for three hysterons, since determining such a bound would require varying all relevant parameters rather than only two. Recent work, however, has begun to address this question \cite{teunisse2025transition}. 

While the methods we have described resolve some of the issues of systems of interacting hysterons and offer a straightforward way of approaching the design of transition graphs, they also become considerably less practical when trying to describe large systems of hysterons. This is primarily due to the computational difficulty of exhaustively solving for the bifurcation points in the governing equations.
Nevertheless, as illustrated in Figures \ref{fig:graphsofgraphs} and \ref{fig:final_graph}, identifying the locations where cusp and crossing bifurcation curves intersect, or higher order bifurcations in general, can guide us to a region in parameter space where many transition graphs can be accessed with only small variations of the parameters around those points. Different continuation algorithms have been developed to find cuspoidal bifurcations of higher codimension \cite{yang2023bifurcation, press2007numerical, kuznetsov2004topological}, but finding bifurcation points of higher codimension that involve crossings of fold curves proves to be more difficult as the number of equations needed to solve for them nearly doubles compared to cusp bifurcations. Future work may focus on developing more efficient numerical techniques to locate these bifurcation points in large systems. Ultimately, bridging the gap between theoretical bifurcation analysis and practical implementation may enable new forms of tunable metamaterials governed by hysteretic behavior.
\section*{Acknowledgments}
We thank Joseph D. Paulsen for helpful discussions. GM and CDG acknowledge funding from the National
Science Foundation through Grant No. EFRI-1935294.
\section*{References}
\bibliographystyle{iopart-num}
\bibliography{bibliography}% 

\end{document}